# Planet Mass and Metallicity: The Exoplanets and Solar System Connection

Mark R. Swain[1] · Yasuhiro Hasegawa[1] · Daniel P. Thorngren[2] · Gaël M. Roudier[1]

**Abstract**
Theoretical studies of giant planet formation suggest that substantial quantities of metals—elements heavier than hydrogen and helium—can be delivered by solid accretion during the envelope-assembly phase. This process of metal enhancement of the envelope is believed to diminish as a function of planet mass, leading to predictions for a mass-metallicity relationship. Supporting evidence for this picture is provided by the abundance of $CH_4$ in solar system giant planets, where $CH_4$ abundance, unlike $H_2O$, is unaffected by condensate cloud formation. However, all of the solar system giants exhibit some evidence for stratification of metals outside of their cores. In this context, two fundamental questions are whether metallicity of giant planets inferred from observations of the outer envelope layers represents the bulk metallicity of these planets, and if not, how are metals distributed within giant planets. Comparing the mass-metallicity relationship for solar system giant planets, inferred from the observed $CH_4$ abundance, with various tracers of metallicity in the exoplanet population, has yielded a range of results. There is evidence of a solar-system-like mass-metallicity trend using bulk density estimates of exoplanet metallicity. However, transit-spectroscopy-based tracers of exoplanet metallicity, which probe only the outer layers of the envelope, are less clear about a mass-metallicity trend and raise the question of whether radial composition gradients exist in some giant exoplanets. The large number of known exoplanets enables statistical characterization of planet properties. We develop a formalism for comparing both the metallicity inferred for the outer envelope and the metallicity inferred using the bulk density and show this combination may offer insights into the broader question of metal stratification within planetary envelopes. Our analysis suggests that future exoplanet observations with JWST and Ariel will be able to shed light on the conditions governing radial composition gradients in exoplanets and, perhaps, provide information about the factors controlling stratification and convection in our solar system gas giants.

## 1 Introduction

The core-accretion model of planet formation (Pollack et al. 1996)—in which a core of solid material is assembled, which then triggers rapid gas accretion of an envelope—provides an

---

     🄢 Springer

explanation for the formation of solar system gas-giant planets that is consistent with planet properties, available material suitable for planet formation, and the lifetime of the protostellar disk. In this scenario, which is also expected to apply to extrasolar gas-giant planets, a planetary core of solid material is believed to be assembled first and then trigger rapid gas accretion to construct an envelope (e.g., Mizuno 1980; Stevenson 1982; Bodenheimer and Pollack 1986; Pollack et al. 1996; Ida and Lin 2004; Benz et al. 2014). During the envelope-accretion phase, various forms of solids (e.g., dust, pebbles, and planetesimals) could be accreted concurrently with gas accretion (e.g., Pollack et al. 1996; D'Angelo and Podolak 2015; Hasegawa et al. 2018; Venturini and Helled 2020), causing the envelope metallicity to deviate from the metallicity (i.e., dust-to-gas ratio) of natal protoplanetary disks. The atmospheric metallicity (the metallicity of the radiative layer of the envelope) controls the efficiency of radiative cooling of energy released by envelope contraction, and thus atmospheric metallicity has an impact on the minimum core mass needed to trigger the onset of rapid gas accretion (e.g., Ikoma et al. 2000; Movshovitz et al. 2010; Mordasini et al. 2014).

The existence of a radial thermal gradient within a protoplanetary disk naturally creates a sequence of ice lines for various volatile materials such that the composition of both the solid and gas components of the protoplanetary disk are a function of radius (e.g., Öberg et al. 2011; Henning and Semenov 2013; Öberg and Bergin 2021). The disk's compositional structure, in turn, was realized to have potentially important consequences for planetary composition, both in terms of the solid material used to assemble the core and the gas-plus-solid material participating in the envelope-accretion process (Öberg et al. 2011; Madhusudhan et al. 2014a; Cridland et al. 2016; Öberg and Bergin 2016).

The core-accretion model in the presence of disk composition structure has also been successfully applied to studies of exoplanet formation (e.g., Mordasini et al. 2012; Thiabaud et al. 2015; Mordasini et al. 2016; Cridland et al. 2016). Important findings from these kinds of studies include: 1) the prediction that there should be a relationship between envelope metallicity and total planet mass, hereinafter referred to as the mass-metallicity relation (Fortney et al. 2013; Mordasini et al. 2016); 2) that the metallicity of the envelope can be altered significantly by solid material accreted with the disk gas (e.g., Venturini et al. 2016; Lozovsky et al. 2017; Stevenson et al. 2022); and 3) that the radial location of the planet at the time of the envelope-building phase may strongly influence the envelope carbon-to-oxygen (C/O) ratio (e.g., Madhusudhan et al. 2014a; Mordasini et al. 2016; Espinoza et al. 2017; Bergin et al. 2023). Theoretical studies also indicate that the size of the solids that participate in the gas-accretion process during the envelope-assembly phase is an important parameter (e.g., D'Angelo et al. 2014; Hasegawa 2022), and if planetesimals are the main agent for metal enrichment of the envelope, less massive (i.e., Neptune-mass) planets may have lower atmospheric metallicities (Fortney et al. 2013) although this may not be the case if pebble accretion dominates solid delivery (Venturini et al. 2016; Ormel et al. 2021).

The above studies therefore strongly suggest that observable signatures in exoplanet atmospheres could be used to probe the process of planet formation through measurement of the mass-metallicity relation, C/O ratios, and other observables. However, in the context of interpreting elemental abundances derived from transit spectroscopy measurements, questions have been raised about the role of heavy element settling and composition gradients (e.g., Mousis et al. 2009, 2011). One of the most critical concerns is that transit spectroscopy samples the outer layers of the atmosphere; if heavy element settling occurs, or if other mechanisms for establishing interior composition gradients exist, transit-derived metallicity estimates may not represent the planet's bulk metallicity, potentially reducing the utility of transit spectroscopy as a direct probe of planet formation processes.

Understanding how metals—that is, any element more massive than helium—are radially distributed within gas-giant planets is one of the major outstanding questions in the field of planetary interiors, for both the planets in our own solar system and exoplanets. Given recent rapid progress in exoplanet measurements, as well as some important advances in planetary studies, it is timely to review relevant work in these areas by evaluating the synergistic relationship between composition studies of both exoplanets and planets in our solar system and identifying what this combined approach might be capable of in terms of yielding insights into planet formation and evolution. This is the goal of our article. The reader is encouraged to refer to e.g., Lissauer and Stevenson (2007), Helled et al. (2014), Guillot et al. (2023) for more general and comprehensive reviews.

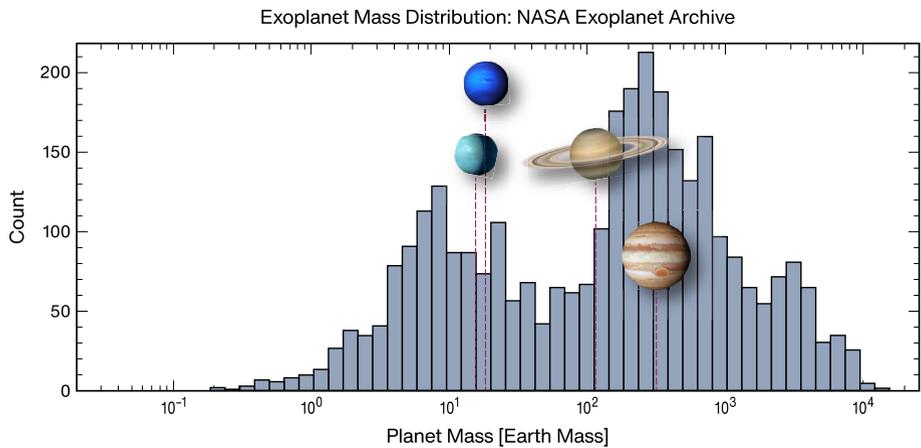

**Fig. 1** When measured by mass, the solar system ice-giant and gas-giant planets provide the best local analogs to the majority of known exoplanets. The symbols for Jupiter, Saturn, Uranus, and Neptune have an accompanying red dashed line illustrating the location of each of the Solar System giants within the distribution of known exoplanet masses

## 2 The Solar System Giant Planets

When compared to their exoplanet cousins, the solar system gas- and ice-giant planets provide accessible environments with the prospect of detailed measurements that would be infeasible in their exoplanet counterparts. Indeed, the solar system giants are in some respects, such as mass and bulk composition, the closest solar system analogs to the majority of the presently known exoplanets (see Fig. 1). For instance, the interior temperatures and pressures of the solar system's gas giants and hot Jupiters are broadly similar (e.g., Guillot 2005), as they are determined mainly by the mass and bulk composition (i.e., hydrogen and helium). Another way in which the solar system gas and ice giants are similar to many exoplanets is that the mean molecular weight of their atmospheres is generally believed to be lower than the available condensibles. The latter point is relevant because moist convection behaves differently in low mean molecular weight atmospheres (e.g., Li and Ingersoll 2015).

The majority of known exoplanets have been discovered by the transit survey method as exemplified by mission such as Kepler and TESS (Borucki 2016; Guerrero et al. 2021). Because the transit geometry is favored by small semi-major axis values, transiting exoplanets typically have higher equilibrium temperatures than planets in the Solar System. Although having higher equilibrium temperatures does not change the planetary bulk composition at the elemental level, it does impact the molecular composition of the planetary atmospheres (Lodders and Fegley 2002). Additionally, processes such as advection may also impact the atmospheric molecular composition (Fortney et al. 2020). While the molecular composition differences between planetary atmospheres are an important subject, our focus here is on the metallicity and not the chemical reservoirs for individual elements.

### 2.1 Overview: Interior Structure

The interior structure of solar system giant planets has been a long-standing topic of inquiry (e.g., Stevenson 1985; Guillot 1999). Direct measurement of gravitational moments through spacecraft trajectory, in situ composition measurements via atmospheric probes, and probing the interior structure with radio transmission measurements or ring seismology, are all techniques that have been used to infer interior structure properties for one or more of the solar system giant planets (e.g., Fuller 2014; Mankovich and Fuller 2021; Guillot et al. 2023, references herein). Of perhaps equal importance is the fact that these inferences about interior structure can be contextualized by comparing them to planetary magnetic field structure and interior heat flow.

As will be discussed further, there are numerous lines of observational evidence for radial (and possibly polar and azimuthal) composition gradients in all of the solar system gas giants, although the nature of the evidence differs among the planets. The presence of these compositional gradients clearly demonstrates the need for improvements on the canonical interior models, wherein planet interiors are represented by two or three well-separated homogeneous layers. In fact, the interior structure of the solar system gas and ice giants remains an area of active investigation, and it is known that uncertainties introduced by the equation of state (EOS) affect both Jupiter core mass estimates and estimates of the deep envelope metallicities for Uranus and Neptune (e.g., Guillot 2005; Fortney and Nettelmann 2010).

### 2.2 Jupiter and Saturn

Jupiter and Saturn are the best solar system analogs for massive hydrogen-dominated exoplanets representing approximately half of the currently known exoplanet population, and evidence for composition gradients in Jupiter and Saturn has been developing for some time. While many of the known exoplanets are in short-period orbits and possess much higher equilibrium temperatures than Jupiter and Saturn, this primarily impacts the outer regions of the envelope; the interior temperatures are more comparable (Guillot 2005).

A key advancement in the understanding of the interiors of Jupiter and Saturn occurred when an improved EOS for hydrogen was used for modeling their interiors (Chabrier et al. 1992) and the need for a heavy-element enhancement in Saturn was found, although the uncertainties associated with deuterium were noted as a limitation to EOS-based modeling (Saumon and Guillot 2004). The earlier picture of a Jupiter interior structure based on three layers (i.e., a core, a metallic hydrogen layer, and a molecular hydrogen layer), with uniform composition within a layer, was already challenged by the incorporation of helium rain-out, although the radial concentration of heavy elements was thought of as approximately uniform (Guillot 2005). It was recognized, however, that if radial composition gradients did exist for metals, there could be an impact on the efficiency of simple convection, as the interior might enter into the doubly diffusive convection regime in the form of layered convection, which has been quantitatively explored (Leconte and Chabrier 2012), where a key parameter is the thickness of the individual layers. If layered convection is operative at some point, the time evolution of the layered convection process is important to consider, but the outcome is unclear; some analysis (Wood et al. 2013) suggests that layers can merge while other studies (Vazan et al. 2018) of Jupiter show that an initially continuous composition gradient can evolve into layered convection. Simulations of plausible formation scenarios for Jupiter find that primordial entropy gradients have the potential to initially suppress convection early in the planet's lifetime (Cumming et al. 2018), which may also play a role in the development of compositional gradients and layering.

The potential connection between core structure and radially extended composition gradients has been brought into sharp focus with recent measurements. The arrival of the Juno probe at Jupiter revealed a gravity field (see Durante et al. 2020) consistent with a diffuse or "fuzzy" core (e.g., Militzer et al. 2022), in which a radially decreasing heavy element enhancement exists throughout the metallic hydrogen region (Wahl et al. 2017), although the detailed density profile is dependent on the assumed EOS (Debras and Chabrier 2019). Studies have found (Helled and Stevenson 2017) that a core lacking a specific, well-defined boundary may be a natural result of planet formation history and could be a common property of giant planets. Some simulations (Moll et al. 2017; Soubiran et al. 2017) find that Jupiter's core, if it initially had a solid, well-defined boundary, would have been eroded rapidly ($\sim 10^6$ years) without the establishment of a "doubly diffusive staircase," resulting in a heavy element radial gradient extending beyond of the original core radius; thus the presence of a core today implies it has been protected by a layered composition gradient. Analysis has established (Debras and Chabrier 2019) that Jupiter's diffuse core could extend to $\sim 0.7$ of the planet radius. Studies have highlighted (Müller et al. 2020) the challenges in the formation of a fuzzy core purely through the planet-formation process. Some investigations suggest (Militzer et al. 2022) that heavy elements from Jupiter's core are distributed to $\sim 0.6$ of the planet's radius and that the core dilution process may still be ongoing. Recent work (Stevenson et al. 2022) has found the metal gradient associated with a mean molecular weight gradient may be established by a hot supersaturated rain-out process during the assembly of Jupiter's envelope; the separate behavior of $Z_{H_2O}$ and $Z_{SiO_2}$ raise the possibility that the C/O ratio might have a radial dependence. Additionally, studies (Stevenson et al. 2022) have also found that the gradients established by the rain-out process were sufficient to suppress convection.

Evidence for composition gradients in the envelope has existed for some time. For example, in a comparison of the interiors of Jupiter and Saturn based on each planet containing three homogeneous layers (core, metallic hydrogen, molecular hydrogen), Guillot (1999) estimated the enrichment of metals in the molecular layer to be 1–6.5 times the solar abundance value for Jupiter and 0.5–12 times the solar abundance value for Saturn. Also, models consistent with inference of the gravitational field detected by the Cassini probe showed Saturn's metallic hydrogen layer extending to approximately 0.5 of the planet's radius (Anderson and Schubert 2007). Similarly, the potential for helium phase separation in Saturn and Jupiter, with the attendant release of latent heat, was first discussed (Stevenson and Salpeter 1977) in the 1970s. Recent theoretical work finds that Saturn has likely formed a deep, helium-rich shell due to helium differentiation (Mankovich and Fortney 2020). Furthermore, a seismology technique based on the rings (Fuller 2014) has been used to determine that Saturn's core is, like Jupiter's, diffuse—this core is estimated to contain $\sim 17$ $M_\oplus$ of rock and ice, is stably stratified, and extends to $\sim 60$% of the planet's radius (Mankovich and Fuller 2021), which is beyond some estimates of the metallic hydrogen region (Mankovich and Fortney 2020). Additionally, the suppression of moist convection, due to the difference in mean molecular weight of the atmosphere and the mean molecular weight of $H_2O$, has been attributed to the basis of periodic storm formation in Saturn's observable atmosphere (Li and Ingersoll 2015).

While the traditional picture of an interior composition gradient is one in which heavy elements are enhanced in the interior, some work has explored the possibility of metallicity enhancement of the outer layers of the envelope (Debras and Chabrier 2019; Howard et al. 2023) of Jupiter. The possibility of metallicity enhancement in the atmosphere and outer portions of Jupiter's envelop is motivated by tension between interior models that explain the gravitational moments measured by Juno (Bolton et al. 2017) and heavy element abundances reported by Galileo (Wong et al. 2004). This possibility of metallicity enhancement in the outer layers of the envelope, including the observable planetary atmospheres, adds a further degree of complexity when considering radial composition gradients.

## 2.3 Uranus and Neptune

Uranus and Neptune, as Fig. 1 shows, are the best solar system analogs, in terms of mass, for a substantial portion of the currently known exoplanet population and recent work (Vazan et al. 2022) indicates the interior temperatures are similar for both short and long period Neptune mass planets. Unlike Jupiter and Saturn, the H/He envelope does not dominate the mass of Neptune and Uranus and contributes less than $\sim 20\%$ to the overall planetary mass (Nettelmann et al. 2013; Helled and Fortney 2020); the majority of the planet mass is comprised mainly of rocks, superionic water (see French et al. 2009), and other "ices". Intriguingly, evidence of composition gradients for Neptune have existed for some time; models assuming uniform composition are inconsistent with gravitational moments (Podolak et al. 1995). Additionally, the magnetic field structure of Neptune and Uranus has been modeled as due to composition gradients that suppress the radial location of the convective region to a relatively thin shell (Stanley and Bloxham 2004). Computational arguments supporting the existence of composition gradients in Neptune and Uranus have also been made (Helled and Fortney 2020). Compared to Jupiter and Saturn, the abundance of carbon and sulphur is enhanced relative to hydrogen in Uranus and Neptune although this may not be the case for nitrogen (see Guillot et al. 2023, and references therein).

A puzzle connected to the current location of Uranus and Neptune is that they would have taken far longer to form at their current locations than the lifetime of the protostellar disk; this suggests that these planets formed closer to the Sun and then underwent migration (Pollack et al. 1996; Frelikh and Murray-Clay 2017; Helled and Bodenheimer 2014, and references therein). Simulations of a range of formation distance scenarios for Uranus and Neptune indicate that the composition of both planets, including the mass of the hydrogen/helium envelope, has high sensitivity (Helled and Bodenheimer 2014) to formation conditions, including the local gas-to-solids ratio in the protostellar disk. Thus, the metallicity of Uranus and Neptune may be more related to their formation location than to their mass.

## 3 Exoplanets

Interest in whether the carbon-based mass-metallicity trend observed in solar system planets was more widely applicable to exoplanets occurred soon after transit spectroscopy measurements began to probe water and methane opacity features in exoplanet atmospheres. An analysis (Mousis et al. 2009) of the early water and methane abundances reported for HD 189733b (Tinetti et al. 2007; Swain et al. 2008) identified the role of heavy element settling as a possible explanation for subsolar abundances; this identification of the possibility of radial composition gradients in exoplanet envelopes remains keenly relevant to studies that searched for observational evidence of an exoplanet mass-metallicity trend.

However, direct comparison between the methane-based metallicity trend observed in solar system gas giants and a methane-based metallicity trend in exoplanets has been problematic to date for three reasons. Firstly, many of the known transiting exoplanets have high equilibrium temperatures that do not favor the formation of methane. Secondly, the most productive instrument to date for transit spectroscopy, the WFC3 instrument on Hubble, provides spectral coverage in a region where water and methane bands can be easily confused. Thirdly, the original detection of methane HD 189733b was disputed (Gibson et al. 2011), provoking questions about how to interpret those measurements. Nonetheless, the importance of comparing the inferred mass-metallicity relationship for Solar System giant planets to that of exoplanets remained widely recognized.

### 3.1 Theory

As discussed in the introduction, extensive theoretical work by a number of groups has examined the question of the deposition of metals during the envelope-assembly phase (Pollack et al. 1996; Ikoma et al. 2000; Alibert et al. 2005b; Klahr and Bodenheimer 2006; Espinoza et al. 2017; Hasegawa et al. 2018; Venturini and Helled 2020; Schneider and Bitsch 2021). Much of this work has focused on the important topic of how the local disk environment at the time of envelope formation may influence the total amount of carbon and oxygen delivered during the formation process (e.g., Öberg et al. 2011; Madhusudhan et al. 2014a; Espinoza et al. 2017); however, some theoretical studies also found and reported a link between planet mass and metallicity (Fortney et al. 2013; Mordasini et al. 2016; Hasegawa et al. 2018) that was qualitatively similar to the mass-to-carbon abundance trend seen in solar system planets. This body of theoretical work, indicating the potential for an exoplanet mass-metallicity relation, has provided a powerful motivator for observational studies to search for trends in exoplanet metallicity and to compare those trends to the solar system.

In addition to finding a general mass-metallicity trend of increasing metallicity with decreasing mass, studies (Fortney et al. 2013) found that the slope of the mass-metallicity trend could reverse at lower planet masses; this was attributed to planetesimals penetrating low-mass envelopes and delivering metals directly to the core. These simulations (Fortney et al. 2013) also found the mass corresponding to a maximum envelope metallicity depended on the characteristic size of the accreting planetesimals: larger planetesimals could penetrate through more massive envelopes and thereby contribute metals to the core instead of to the envelope. For planets on close-in orbits, such low-mass planets (especially those with metal-poor envelopes) are vulnerable to mass-loss (e.g., Lopez et al. 2012), potentially complicating the inference of primordial envelope metallicity in planets subject to significant envelope loss.

### 3.2 Observations

In one of the first observation tests of a mass-metallicity relation, transit spectroscopy measurements obtained with Hubble's Wide Field Camera 3 (WFC3) G141 grism, (Kreidberg et al. 2014b) were used to make a direct comparison between the transit-derived metallicity of WASP-43 b and the Carbon-mass relationship in the Solar System giant planets. As the sample of exoplanets observed with WFC3/G141 grism grew, additional estimates of the exoplanet mass-metallicity relation were reported, leveraging the relatively strong water band probed by Hubble WFC3. An early result (Madhusudhan et al. 2014b) reported subsolar water abundance of three hot Jupiters. Subsequently, another study (Barstow et al. 2017) found a trend of subsolar water abundance in a sample of 10 hot Jupiters. Additional studies (Wakeford et al. 2017a, 2018; Wakeford and Dalba 2020) using Hubble WFC3 observations found mass-metallicity trends qualitatively similar to the solar system, while other work (Pinhas et al. 2019), using metallicity derived from $H_2O$ abundance, found no evidence for a mass-metallicity relationship (although a they did find a systematic depletion of $H_2O$ relative to Solar abundance), a finding that was inconsistent with the solar system. Another recent study (Edwards et al. 2023) found no evidence for a mass-metallicity trend from Hubble WFC3 observations. Taken together, there is a wide range of findings in these studies, from solar system–like mass-metallicity trends (e.g., Wakeford and Dalba 2020) to no evidence of a mass-metallicity relationship (e.g., Edwards et al. 2023); however, a common denominator for these studies was the finding of substantial metallicity scatter in the exoplanet population, suggesting that exoplanets have diverse properties.

Metallicity estimates based on Hubble WFC3 measurements have three potentially important limitations. Firstly, many of the WFC3-based studies used the abundance of water as a proxy for metallicity (invoking either explicitly or implicitly the existence of approximately solar C/O values and thermal equilibrium chemistry), which substantially complicates the comparison to the methane-based mass-metallicity relation for Solar System giants. Secondly, many of the Hubble WFC3-based exoplanet metallicity estimates were based on samples that were not uniformly processed, relying instead on aggregating metallicity findings from multiple teams and data reduction methods. Thirdly, the absence of optical spectral coverage in the near-infared WFC3 measurements reduces sensitivity to aerosols (discussed in more detail in the following section). Recognizing the potential limitations of the existing WFC3-based estimates of metallicity, other measures of metallicity, based on sodium and potassium abundance, were reported (Welbanks et al. 2019) and compared to Hubble WFC3-based measurements; a key finding of this study was the mass-metallicity relation inferred using the abundances of sodium and potassium differed from the mass-metallicity relation inferred using $H_2O$ abundance, indicating a potential sensitivity of the transit-derived mass-metallicity relationship to the species, whether atomic or molecular, being employed as a basis of the metallicity estimate.

The process of estimating metallicity from transit spectra requires modeling the observations; the assumptions that go into this process can vary widely. Some of the ways in which retrieval assumptions can vary include, but are not restricted to, the representation of clouds and aerosols, temperature profile, vertical mixing ratio of absorbing species, and the kinds of chemical/physical processes incorporated in modeling atmospheric composition. For example, of the kinds of chemical/physical processes that have been used in modeling exoplanet spectra include free retrievals, where a prescribed set of opacity sources are allowed to independently, and retrievals where the opacity sources are linked by models such as thermal equilibrium chemistry, photochemistry, and advection. A study of the kinds of retrieval assumptions that can be constrained by transit spectra data (Fisher and Heng 2018) found that relatively simple assumptions, such as an isothermal atmospheres, are typically reasonable for HST measurements but that more complex models may be needed for JWST observations. There is also some evidence that the details of cloud models do not qualitatively impact the identification of molecular species (Barstow 2020). However, when retrieval codes are intended to implement the same atmospheric modeling assumptions, there are questions about the level of mutual agreement and efforts have been made to cross-validate retrievals under controlled circumstances (Barstow et al. 2020). A review of exoplanet retrieval methods is beyond the scope of this manuscript but the possibility for atmospheric modeling assumptions to influence estimates of atmospheric metallicity is clear; this area would benefit from continued investigation.

Although less numerous, in terms of the planets that have been studied, than space-based transit spectroscopy measurements, the promising approach of high spectral resolution ground-based observations can also probe the metallicity of exoplanet atmospheres. An example of this promising class of observations is recent work identifying WASP-77Ab as having a sub-Solar metallicity atmosphere (Line et al. 2021) with uncertainties that are much smaller than typical metallicity uncertainties determined from Hubble transit spectra. The demonstrated ability of high spectral resolution observations to detect numerous atomic species in hot Jupiter atmospheres (Pelletier et al. 2021, 2023; Gandhi et al. 2023) suggests that the use of multiple elemental ratios would be a fruitful future line of investigation that goes beyond the discussion of metallicity we explore in this manuscript.

### 3.3 The Role of Aerosols

It was recognized from the outset (Charbonneau et al. 2002) that the presence of aerosols, whether in the form of condensible clouds or photochemical hazes, could reduce the amplitude of features in the transit spectrum, and therefore bias estimates of metallicity, unless aerosols were specifically modeled as part of the spectral retrieval process (e.g., Kreidberg et al. 2014b). Although early retrieval codes did not always include explicit modeling of aerosols, it is now common practice for retrieval codes used for interpreting exoplanet transit spectroscopy measurements to include a prescription for modeling clouds and hazes (e.g., Line and Yung 2013; Waldmann et al. 2015; Irwin et al. 2008; Barstow et al. 2017; Roudier et al. 2021; Harrington et al. 2022).

Some transit spectra show simultaneous evidence for truncation of spectral modulation in the near-infrared through a "bottom cloud" and a Rayleigh-like scattering slope attributed to haze at shorter wavelengths (Wakeford and Dalba 2020). Transit spectra of some planets show evidence of aerosols at ∼microbar pressures (Estrela et al. 2021), which is approximately consistent with the ∼0.1 microbar pressure identified (Lavvas and Koskinen 2017) as the key region for haze monomer formation. Microphysical modeling suggests (Lavvas and Koskinen 2017) that monomer formation is followed by a sedimentation process that can distribute an aerosol throughout much of the observable atmospheric column, and a recent observational study (Estrela et al. 2022) reported evidence that it is common for transiting exoplanets to have an aerosol component distributed from ∼millibar to ∼microbar pressures.

Since the signature of Raleigh scattering is stronger in the optical than in the near-infrared, efforts have been made to incorporate optical data, using the Hubble STIS instrument or ground-based measurements, in the analysis of Hubble WFC3 spectra (Pinhas et al. 2019). In principal, incorporating optical measurements allows better constraints on the haze/cloud model parameters and, depending on spectral resolution and coverage, access to additional species, such as sodium, for tracing metallicity. However, cooler exoplanet host stars can be quite faint in the visible, making high-precision transit spectroscopy measurements difficult in these cases. There are also challenges associated with combining optical and near-infrared data sets; these include modeling the inhomogeneous nature of the stellar photosphere, an effect that is stronger in the visible wavelengths, (Rackham et al. 2018; Iyer and Line 2020). Nonetheless, obtaining transit spectral coverage in the optical remains a scientifically important objective when the measurements are feasible.

Aerosols may have impacts beyond merely obscuring, or partially obscuring, the spectral signatures of atomic and molecular species. The potential sequestration of carbon in polycyclic aromatic hydrocarbons and soots was identified (Mousis et al. 2011) as a possible source of gas-phase depletion of carbon. This concept of sequestration can be generalized to aerosols, whether formed by photochemical or condensation processes. Theoretical studies of likely condensible species for exoplanet atmospheres showed a range of candidate species for a wide range of atmospheric temperatures and pressures (Morley et al. 2013), and observational studies have established that transiting exoplanets, as a class of objects, frequently contain sufficient clouds and haze to impact the transmission spectra (Sing et al. 2016; Stevenson 2016; Iyer et al. 2016; Gao et al. 2021; Estrela et al. 2022). These clouds could sequester metallicity tracers and the potential for sequestering species such as sodium in clouds was discussed as a possible interpretation for the sub-Solar sodium abundance reported in HD 209458 b (Charbonneau et al. 2002).

We still lack observational evidence pointing to a specific origin scenario for hazes in the atmospheric column. While the possibility of photochemical origin exists, a condensation-based origin is difficult to rule out at the present time; however, the evidence for significant photochemistry in exoplanet atmospheres is accumulating (Roudier et al. 2021; Rustamkulov et al. 2023), and the area remains one of active investigation.

### 3.4 Bulk Metallicity

In contrast to the numerous transit spectroscopy studies that have searched for a mass-metallicity relationship, the bulk density has been used to estimate the metal content in exoplanets (Guillot et al. 2006; Miller and Fortney 2011). The most recent study using this approach uniformly analyzed a sample of 47 planets (Thorngren et al. 2016); to minimize the potential bias based on hot-Jupiter inflation, that study was restricted to planets with $T_{eq} < 1000$ K as previous work suggested that giant planets at those temperatures do not show evidence for being inflated (Miller and Fortney 2011). The bulk density-derived metallicity (Thorngren et al. 2016) analysis found a mass-metallicity relationship that is qualitatively similar, although not identical, to the mass-carbon relationship for Solar System giant planets; however, the bulk density-derived metallicity (Thorngren et al. 2016) had substantial scatter around the underlying trend, and the standard deviation of mass-metallicity model residuals is 7.6 in units of planet metallicity divided by stellar metallicity ($Z_{planet}/Z_{star}$). This scatter of the data around the trend line is larger than the median $Z_{planet}/Z_{star}$ uncertainty of 4.6, suggesting that this scatter is an intrinsic property of the exoplanet population. This scatter is likely due to the range of typical gas-to-solid ratios encountered during envelope assembly, and a scatter around a mass-metallicity trend was also found in theoretical studies by Mordasini et al. (2016).

Although atmospheric spectroscopy and bulk metallicity models both measure the composition of planets, they are not equivalent. Most significantly, the bulk metallicity may be larger than the atmospheric metallicity due to the presence of a composition gradient or a metal-rich planetary core. The bulk metallicity determined by structure models cannot distinguish between types of metals, having relied on assumptions about their ratios—typically iron and silicates versus volatile species such as water, methane, and ammonia. On the other hand, while atmospheric metallicities can sometimes distinguish between certain species, not all molecules or atomic species are amenable to spectroscopy at all temperatures. For example, silicates can only be seen in the atmospheres of very hot Jupiters (e.g., Lothringer et al. 2022); and although neon is thought to be relatively abundant in the sun (Asplund et al. 2009), it lacks any detectable lines at planetary temperatures so is not directly detectable by transmission spectroscopy. As such, comparisons between bulk metallicity, atmospheric abundances, and a planet's formation history should be done thoughtfully.

Even relatively simple comparisons have pitfalls to avoid—consider the conversion from mass ratio to number ratio discussed in Thorngren and Fortney (2019):

$$Z{:}H = \frac{1 + Y/X}{(Z^{-1} - 1)(\mu_Z/\mu_H)}. \tag{1}$$

In the above equation, using the solar helium abundance and metallicity (Asplund et al. 2009) under planetary conditions yields a number ratio of 0.002—roughly twice the number ratio of the sun. This is because, unlike the sun, planets contain molecular hydrogen, halving the baseline number of hydrogen. It is important not to interpret the conversion from mass ratio to a number ratio as a metal enhancement from planet formation, when it results solely from lowering the temperature of the gas. Atmospheric composition assumptions may also produce incorrect results by using the wrong mean molecular mass $\mu_Z$: a water-dominated atmosphere will have $\mu_Z \approx 18$, whereas a $CO_2$-dominated atmosphere has $\mu_Z \approx 44$.

### 3.5 Recent Observational Developments

The exoplanet mass-metallicity relationship continues to be a topic of interest for the recently launched James Webb Space Telescope (JWST), with observations of three planets, WASP-39 b, HD 149026 b, and WASP-80 b to date, with reported metallicity determinations (Rustamkulov et al. 2023; Feinstein et al. 2023; Alderson et al. 2023; Ahrer et al. 2023; Bean et al. 2023; Bell et al. 2023). Four different instrument modes were used to measure WASP-39 b, and while the metallicity findings from these observations were broadly consistent in finding super-solar metallicity, the detailed metallicity values determined with individual instrument modes are not the same, suggesting this as an area for future investigation; we also note the finding (Welbanks et al. 2019) that metallicity can depend on the measurement method. Analysis of the measurements of WASP-80 b (Bell et al. 2023) atmosphere with a metallicity enhancement of between 3–10× with respect to Solar, which is consistent with the bulk-metallicity determination for this planet (Thorngren and Fortney 2019), suggesting that this planet may not support strong composition gradients. Taken together, the JWST measurements of the WASP-39 b, HD 149026 b, and WASP-80 b appear consistent with the finding by Thorngren et al. (2016) of substantial scatter in the trend line of the mass-metallicity relationship. Recent JWST observations of the sub-Neptune K2-18 b indicate metallicity enhancement above solar values (Madhusudhan et al. 2023). Additionally, the JWST observations detected methane in both K2-198 b and WASP-80 b, opening the possibility of direct comparison, using the same atmospheric metallicity indicating species, of the mass-metallicity relationship between Solar System giants and exoplanets.

As the body of observations that constrain exoplanet atmospheric metallicity continues to grow, comparing the density-derived mass-metallicity trend, which probes the bulk metal content, and the transit-derived mass-metallicity trend, which probes the abundance of metals in the outer envelope, has the capability, in principle, to reveal if composition gradients exist. This type of work is urgently needed, as how the metals in the envelope are radially distributed is a matter of great importance to the interpretation of the many ongoing planet spectroscopy programs. An example comparison is provided in the following section.

## 4 Discussion

As described in Sect. 2, evidence for composition gradients exists in all of the solar system giants. Whether these gradients are produced by gravitational settling (Stevenson 1985), par-tially dissolving the core (Wilson and Militzer 2012), the planet formation process (Stevenson et al. 2022), or some combination of these processes, is not yet clear; however, given the abundance of evidence for composition gradients in the Solar System giants, it seems reasonable to assume composition gradients are likely to exist for some exoplanets.

If composition gradients in exoplanets exist, the efficiency of convection can be substantially reduced (e.g., Leconte and Chabrier 2012; Helled and Stevenson 2017), impacting both magnetic field structure (Stanley and Bloxham 2004) and generation (Mankovich and Fortney 2020), and, in sub-Neptunes, the heat flow available to drive envelope mass loss (Misener and Schlichting 2022). The atmospheric metallicity may also grow with time as interior layers mix with the outermost layer that feeds the atmosphere; for example, Vazan et al. (2018) construct a layered model of Jupiter in which the outer layer's $Z$ nearly doubles to 0.08 over its lifetime.

### 4.1 Presence of the Exoplanet Mass-Metallicity Relation?

At the present time, the observational evidence for an exoplanet atmosphere mass-metallicity trend is mixed with studies using Hubble transit spectra (see the previous discussion), yielding a range of results, from a trend similar to what is seen in Solar System gas giants to no evidence for a trend. The limited spectral coverage of the currently available Hubble observations may be an important factor in limiting the current results (Welbanks and Madhusudhan 2019; Welbanks et al. 2019; Pinhas et al. 2019). The dominant atomic species in the metallicity enhancement are expected to be O, C, and N, under the assumption that the relative abundance of individual metals follows solar trends. Thus, to determine the metallicity via the transit spectroscopy method, we need good sensitivity to the major O, C, and N molecular species ($H_2O$, CO, $CH_4$, $CO_2$, HCN, $NH_3$) that may be present (Moses et al. 2013) in a hydrogen-dominated atmosphere; JWST and the future Ariel mission have the spectral coverage for good sensitivity to these molecules. However, currently, there is not clear agreement in the published mass-metallicity trend results obtained with Hubble transit observations.

### 4.2 Link Between Bulk and Atmospheric Metallicities

Bulk density studies of gas-giant exoplanets reveal that the majority of metals are not contained in the core and instead reside in the envelope (Thorngren et al. 2016), and that there is a solar system–like trend for the mass-metallicity relationship when metallicity is estimated using the planet's bulk density. If the planet envelopes are well-mixed, we would expect the transit-derived mass-metallicity relationship to have a form similar to the bulk density-derived mass-metallicity relationship (cf. Wakeford and Dalba 2020). However, if exoplanet envelopes are not well-mixed and composition gradients exist, we would expect a mass-metallicity trend that is weaker or perhaps nonexistent (cf. Edwards et al. 2023). Thus, comparing the density-derived mass-metallicity trend, which probes the bulk metal content, and the transit-derived mass-metallicity trend, which probes the abundance of metals in the outer envelope, has the capability, in principle, to reveal if composition gradients exist.

We propose two metrics to explore this question, one metric being a comparison power law exponent values used to model the mass-metallicity trends and one being a comparison of the relative scatter of the data around the models. Since bulk density-derived and transit-derived studies may not include the same planets and may sample the range of exoplanet masses differently, it is advantageous to track scatter around the model in terms of fractional error defined here as (data − model)/data. Thus our metrics are,

$$\alpha \equiv \frac{m_T}{m_B}, \tag{2}$$

where $m_T$ and $m_B$ are the slopes of the transit-derived and bulk density-derived mass-metallicity relationship in log space, and

$$\beta \equiv \frac{\sigma_T}{\sigma_B}, \tag{3}$$

where $\sigma_T$ is the standard deviation of the relative error for the transit-derived mass-metallicity relationship and $\sigma_B$ is the same quantity for the bulk density-derived mass-metallicity relationship. Values of $\alpha \sim 1$ would indicate a well-mixed condition or an absence of composition gradients, while values approaching zero would indicate the presence of composition gradients. Values of $\beta > 1$ could potentially indicate different amounts of depletion of metals in the outer layers of the atmosphere relative to the bulk density composition and thus could be useful as a diagnostic of the variation in degree and extent of the processes that may establish, or reduce, composition gradients. The possibility of metallicity enhancement in the outer layers of the envelope (discussed in Sect. 2.2), could influence $\beta$ values. Assuming that such an enhancement was significant, relative to deeper interior composition gradients, and occurred in some fraction of giant planets, values of $\beta$ would tend to be larger than would otherwise be the case. However, if the metallicity enhancement of the outer layers of the envelope is small compared to the overall metallicity gradient, as some models of Jupiter suggest (Debras and Chabrier 2019), the overall composition difference between the deeper interior regions and the atmosphere might still be detectable.

### 4.3 Density-Derived Metallicity vs Transit-Derived Metallicity

As described above, it is crucial to compare the density-derived metallicity ($Z_{p,\text{bulk}}$), which probes the bulk metal content, with the transit-derived metallicity ($Z_{p,\text{env}}$), which probes the abundance of metals in the outer envelope. However, the former metallicity is obtained from mass estimates, while the latter is generally from envelope heavy element abundance. It is ideal to use the same measure (mass versus abundance) to define both metallicities. In Appendix A, we describe how the abundance-based, transit-derived metallicity is converted to the mass-based one. We here provide an example population-level comparison between density-derived metallicity and transit-derived metallicity using previously published studies. It should be noted that the following exercise is only for illustrative purposes since there are significant caveats when using the HST-WFC3-only results, as described above.

Figure 2 shows our results of $Z_{p,\text{env}}$. In this plot, we have used the data of Edwards et al. (2023) for exoplanets; the definition of $Z_P$ is unclear in their paper (see their Table 15), and we have here assumed that $Z_P \equiv$ X/H, which is the standard outcome of retrieval analysis based on thermal equilibrium chemistry (see Appendix A). For comparison purpose, the data of $Z_{p,\text{bulk}}$ for exoplanets are taken from Thorngren et al. (2016) and plotted in the figure. Also, solar system planets (i.e., Jupiter, Saturn, Uranus and Neptune) are included; $Z_{p,\text{env}}$ is computed, using the C/H data (Guillot et al. 2023, references herein), and $Z_{p,\text{bulk}}$ is estimated, adopting the results of interior modeling (Saumon and Guillot 2004; Helled et al. 2011; Wahl et al. 2017). Both $Z_{p,\text{env}}$ and $Z_{p,\text{bulk}}$ are normalized by the host stellar metallicity, which is defined as, following Thorngren et al. (2016),

$$Z_s = 0.014 \times 10^{[\text{Fe/H}]}. \tag{4}$$

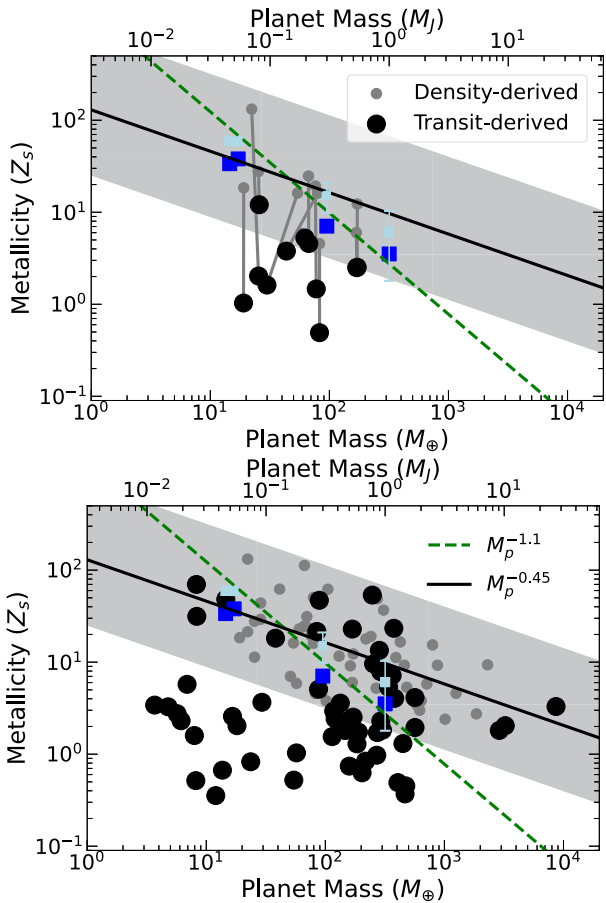

**Fig. 2** Transit-derived ($Z_{p,\text{env}}$) and density-derived ($Z_{p,\text{bulk}}$) metallicities of solar and extrasolar planets. The former metallicities are computed using the approach described in Appendix A. The data is taken from Edwards et al. (2023) for exoplanets (the black dots), while for solar system planets (the blue squares), the C/H ratio is used (Guillot et al. 2023). For the latter metallicities ($Z_{p,\text{bulk}}$), the results of Thorngren et al. (2016) are adopted for exoplanets (the gray dots), while for solar system planets, the results of interior modeling are used (Saumon and Guillot 2004; Helled et al. 2011; Wahl et al. 2017, the light blue squares). For comparison purpose, the fitting results of Kreidberg et al. (2014b) and Thorngren et al. (2016) are denoted by the green dashed and the black solid lines, respectively. The top panel summarizes exoplanets that have both $Z_{p,\text{env}}$ and $Z_{p,\text{bulk}}$, which are constrained by previous studies. The bottom panel contains exoplanets that have either $Z_{p,\text{env}}$ or $Z_{p,\text{bulk}}$. Transit-derived metallicities of exoplanets exhibit a diverse distribution, implying the importance of subsequent evolution processes in exoplanet atmospheres

Our calculations show that for Jupiter, $Z_{p,\text{bulk}} \simeq Z_{p,\text{env}}$ within the uncertainties (see the blue and light blue squares). This implies that heavy elements could be entirely well-mixed within Jupiter and the presence of a well-defined core would not be fully justified. Clearly, this is not consistent with the current picture and better measurements are needed even for Jupiter. One potential explanation may be that $Z_{p,\text{env}}$ is derived from the C/H data; Carbon is not necessarily well-mixed within Jupiter, but its partial mixing might be significant enough. In other words, Carbon may not be a good tracer of the interior structure of Jupiter. On the other hand, other solar system planets and all the exoplanets show deviations between $Z_{p,\text{bulk}}$ and $Z_{p,\text{env}}$, suggesting that heavy elements are not entirely well mixed within these planets (the top panel). Note that mass estimates of Kepler 9 b and c are controversial in the literature (e.g., Borsato et al. 2019), and different values are used in Thorngren et al. (2016) and Edwards et al. (2023) (see the gray thin lines connecting the grey and black dots on the top panel of Fig. 2).

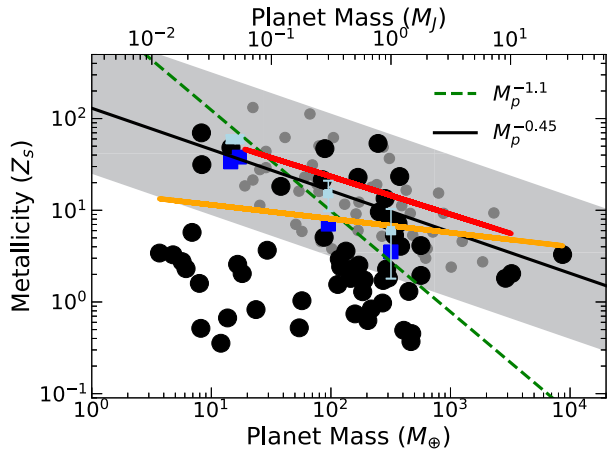

**Fig. 3** Transit-derived ($Z_{p,\text{env}}$) and density-derived ($Z_{p,\text{bulk}}$) metallicities of solar and extrasolar planets, as shown in Fig. 2. Our best fits are plotted by the orange and red solid lines for $Z_{p,\text{env}}$ and $Z_{p,\text{bulk}}$, respectively

Figure 2 (bottom) also shows that $Z_{p,\text{env}}$ has a more diverse distribution than $Z_{p,\text{bulk}}$. This may suggest that the importance of subsequent atmospheric evolution is different for different exoplanets; one potential key question for the future work may be what physical process(es) and/or parameter(s) would control the efficiency and the outcome of subsequent atmospheric evolution.

In order to conduct a more quantitative analysis, we apply the metrics proposed in Sect. 4.2 to the data shown on the bottom panel of Fig. 2. We find that our best fits lead to $m_T \simeq -0.15$, $m_B \simeq -0.41$, $\sigma_T \simeq 6.2$, and $\sigma_B \simeq 1.4$. as shown in Fig. 3. Consequently, our metrics ($\alpha$ and $\beta$) become (see equations (2) and (3))

$$\alpha \simeq 0.37, \quad (5)$$

$$\beta \simeq 4.4.$$

This tentative result suggests that composition gradients are common for currently observed exoplanets since $\alpha < 1$ and bulk metallicities are less diverse than envelope metallicities since $\beta > 1$.

It would be critical to examine if the above two values would behave similarly or differently when more data are included and/or when the whole samples would be divided into sub-groups in terms of planet properties (e.g., orbital periods, multiplicity, age, etc).

### 4.4 Implications for the Heavy-Element Distribution Within Planets

The estimate of $Z_{p,\text{env}}$ and $Z_{p,\text{bulk}}$ enables one to constrain the heavy-element distribution within planets. We here provide preliminary calculations to estimate the heavy element radial distribution.

Planets that are composed of gas (i.e., hydrogen and helium) and solids (i.e., ice and rocks) have the following mass ($M_p$):

$$M_p = M_{XY} + M_Z, \quad (6)$$

where $M_{XY}$ and $M_Z$ are the total gas and solid masses, respectively. For such planets, the density-derived metallicity ($Z_{p,\text{bulk}}$) is defined as

$$Z_{p,\text{bulk}} \equiv \frac{M_Z}{M_p}. \tag{7}$$

On the other hand, the observed transit-derived metallicity *by mass* ($Z_{p,\text{env}}$) should be written as

$$Z_{p,\text{env}} \equiv \frac{M_{Z,\text{env}}^{\text{ob}}}{M_{XY}^{\text{ob}} + M_{Z,\text{env}}^{\text{ob}}}, \tag{8}$$

where $M_{Z,\text{env}}^{\text{ob}}$ is the heavy element mass distributing in planet envelopes that amount to $M_{XY}^{\text{ob}}$ and are probed by transit observations.

Transit spectroscopy can probe only the outer (thin) layer of planet envelopes (i.e., $M_{XY}^{\text{ob}} \neq M_{XY}$), and when planets have radial compositional gradients, $Z_{p,\text{env}} M_{XY} \neq M_Z$. However, if well-mixed assumptions would be adopted, one can readily estimate the heavy element mass ($M_{Z,\text{env}}$) distributing in planetary envelopes. In the following, we consider three limiting cases and compute the value of $M_{Z,\text{env}}$, using $Z_{p,\text{env}}$ with some assumptions.

*The entirely well-mixed case.* This assumes that all the metals that otherwise constitute planetary cores and metals in envelopes would be *entirely* well-mixed with the envelope gas, that is,

$$Z_{p,\text{env}} \simeq \frac{M_{Z,\text{env}}}{M_p}. \tag{9}$$

In other words, there are no well-defined cores. For this case, $M_{Z,\text{env}}$ can be computed as (using equation (9))

$$M_{Z,\text{env}}^{\text{ent}} \simeq M_Z \simeq Z_{p,\text{env}} M_p. \tag{10}$$

Equivalently,

$$Z_{p,\text{env}} = Z_{p,\text{bulk}}. \tag{11}$$

This indicates that the entirely well-mixed case leads to the observed transit-derived metallicity by mass becoming equal to the density-derived metallicity. As discussed above, Jupiter can hold this case within the current error bars (see Fig. 2). Care is needed when this case is applied to Neptune-like planets; as expected from Uranus and Neptune, such planets should have well-defined cores, and hence the resulting estimate should be viewed as an upper limit.

*The globally well-mixed case.* This assumes that $M_{Z,\text{env}}$ would uniformly distribute *globally* within whole envelopes. This does not necessarily preclude the possibility that metals may concentrate at the central region of planets due to the presence of planetary cores and/or subsequent sedimentation of metals. The latter would be unavoidable when refractory materials are accreted. Therefore, $M_Z$ can be decomposed into

$$M_Z = M_{Z,\text{cen}} + M_{Z,\text{env}}, \tag{12}$$

where $M_{Z,\text{cen}}$ represents the heavy element mass located in the central region of planets. This decomposition ends up with three unknowns (i.e., $M_{XY}$, $M_{Z,\text{cen}}$, and $M_{Z,\text{env}}$), but two constraints (i.e., equations (6) and (8)) unless $Z_{p,\text{bulk}}$ (or $M_Z$) is additionally constrained. If the additional constraint (i.e., $Z_{p,\text{bulk}}$ or $M_Z$) is available, then

$$Z_{p,\text{env}} \simeq \frac{M_{Z,\text{env}}}{M_{XY} + M_{Z,\text{env}}} \simeq \frac{M_{Z,\text{env}}}{M_p - M_Z + M_{Z,\text{env}}}. \tag{13}$$

Consequently, $M_{Z,\text{env}}$ can be written as (using equation (13))

$$M_{Z,\text{env}}^{\text{glo}} \simeq \frac{Z_{p,\text{env}}}{1 - Z_{p,\text{env}}}(M_p - M_Z). \tag{14}$$

*The globally well-mixed case with the gas giant assumption.* The additional constraint (i.e., $Z_{p,\text{bulk}}$ or $M_Z$) is not always available for exoplanets. If planets are classified as gas giants, then it is reasonable to assume that $M_p \simeq M_{XY}$. For this case, equation (13) can be written as

$$Z_{p,\text{env}} \simeq \frac{M_{Z,\text{env}}}{M_{XY} + M_{Z,\text{env}}} \simeq \frac{M_{Z,\text{env}}}{M_p + M_{Z,\text{env}}}. \tag{15}$$

Eventually, $M_{Z,\text{env}}$ can be given as (using equation (15))

$$M_{Z,\text{env}}^{\text{glo,GG}} \simeq \frac{Z_{p,\text{env}}}{1 - Z_{p,\text{env}}} M_p, \tag{16}$$

where GG stands for gas giants. As with the entirely well-mixed case, this case would provide an upper limit of $M_{Z,\text{env}}$ for Neptune-like planets.

In summary, one can readily compute $M_{Z,\text{env}}$ for the above three limiting cases, if $M_p$, $Z_{p,\text{env}}$, (and $Z_{p,\text{bulk}}$) are provided.

We now apply the above three cases to exoplanets. Figure 4 shows our results of $M_{Z,\text{env}}^{\text{ent}}$, $M_{Z,\text{env}}^{\text{glo}}$, and $M_{Z,\text{env}}^{\text{glo,GG}}$. As done in Fig. 2, the data from previous studies are used. One immediately notices that $M_{Z,\text{env}}^{\text{ent}} \simeq M_{Z,\text{env}}^{\text{glo,GG}}$ for most exoplanets (see the yellow and black dots on the both panels). This is direct reflection that $Z_{p,\text{env}} \ll 1$ (see equations (10) and (16)); deviations occur when $Z_{p,\text{env}}/Z_s$ takes an order of 100 (see Fig. 2). It is also obvious that the values of $M_{Z,\text{env}}^{\text{ent}}$ and $M_{Z,\text{env}}^{\text{glo,GG}}$ are lower than the green dash line of $M_Z = M_p$, except for HD 97658 b (the right panel); such a planet has the highest value of X/H in Edwards et al. (2023) and is less massive. Therefore, the gas giant assumption (i.e., $M_p \simeq M_{XY}$) is clearly unjustified for most of massive planets (i.e., $M_p > 100 M_\oplus$). The outcome that $M_{Z,\text{env}}^{\text{ent}} \simeq M_{Z,\text{env}}^{\text{glo,GG}} < M_p$ is one necessary confirmation that retrieved analysis is reasonable.

Our calculations also show that the globally well-mixed case leads to the lowest values of $M_{Z,\text{env}}$ (see the red dots on the left panel). This indicates that metal sedimentation should have occurred for all of these exoplanets, and heavy elements would concentrate in the central region currently; that is why $Z_{p,\text{env}} < Z_{p,\text{bulk}}$ as shown in Fig. 2.

It is interesting that $M_{Z,\text{env}}^{\text{glo}} \simeq M_{Z,\text{env}}^{\text{ent}} \simeq M_{Z,\text{env}}^{\text{glo,GG}}$ for massive planets ($M_p \gtrsim 100 M_\oplus$) while $M_{Z,\text{env}}^{\text{glo}} < M_{Z,\text{env}}^{\text{ent}} \simeq M_{Z,\text{env}}^{\text{glo,GG}}$ for less massive planets. Mathematically, this is explained by that massive planets achieve $M_{XY} \simeq M_p$ (equivalently, $M_{XY} \gtrsim M_Z$, see equation (14)). On the other hand, less massive planets have high values of $M_Z$ (i.e., $M_{XY} \lesssim M_Z$). Consequently, $M_{Z,\text{env}}$ decreases (see equation (14)); as described above, both the entirely well-mixed case and the globally well-mixed case with the gas giant assumption provide an upper limit. This planet mass dependence may also be relevant to low-mass planets having tenuous

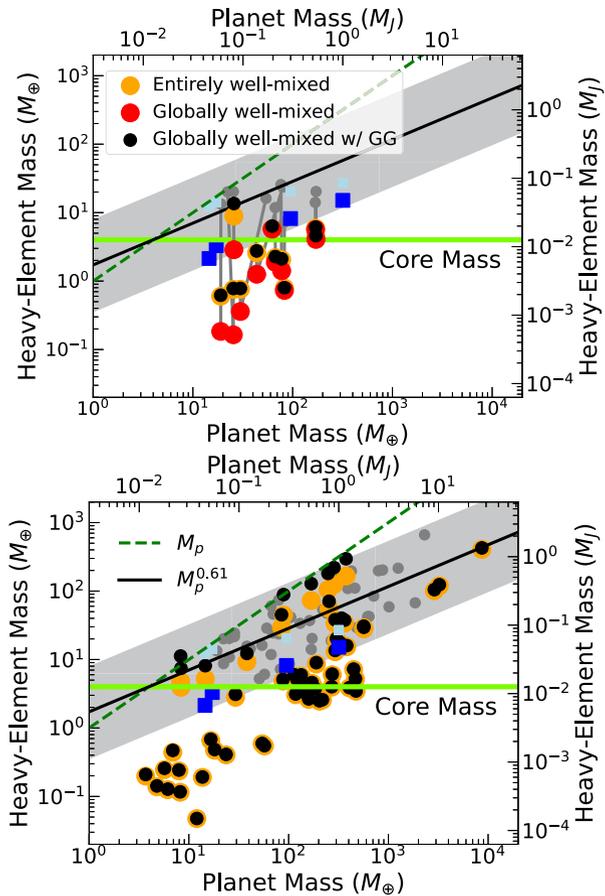

**Fig. 4** Heavy element mass distributing in planetary envelopes ($M_{Z,\text{env}}$) and the total heavy element mass ($M_Z$) of solar and extrasolar planets. As in Fig. 2, the data are taken from previous studies. For $M_{Z,\text{env}}$ of exoplanets, the entirely and globally well-mixed cases and the globally well-mixed case with the gas giant assumption are denoted by the orange, red, and black dots, respectively. For solar system planets, the globally well-mixed case is considered to compute $M_{Z,\text{env}}$ (the blue squares). For comparison purposes, $M_Z$ of solar system planets and exoplanets are shown by the light blue squares and the gray dots, respectively. Also, the fitting result of Thorngren et al. (2016) and the line that $M_Z = M_p$ are denoted by the black solid and green dashed lines, respectively. A predicted core mass is denoted by the light green (Hasegawa et al. 2018). On the top panel, exoplanets that have constraints on both $Z_{p,\text{env}}$ and $Z_{p,\text{bulk}}$ are summarized, while on the bottom panel, those that have a constraint on either of them are plotted

envelopes, so that metals do not dissolve into the envelopes efficiently during the formation stage or that metal sedimentation is efficient in the subsequent evolution stage.

Our efforts thus suggest that for massive planets, both the entirely well-mixed case and the globally well-mixed case with the gas giant assumption can reproduce the globally well-mixed case well unless $Z_{p,\text{env}}/Z_s$ is an order of 100 or higher. For less massive planets, both the entirely well-mixed case and the globally well-mixed case with the gas giant assumption provide the conservative upper limit for $M_{Z,\text{env}}$. Furthermore, $M_{Z,\text{env}}$ tends to be lower than $M_Z$, implying that heavy elements could distribute more in the central region of planets.

We also focus on planets that have both constraints of $Z_{p,\text{bulk}}$ and $Z_{p,\text{env}}$ and plot their mass budget and metallicity profiles in Figs. 5, 6, 7, and 8, respectively. The globally well-mixed case is considered. Our results suggest that heavy elements in envelopes tend to increase with increasing planet mass (Fig. 5). However, most of these metals distribute mainly in the central region of planets currently. This trend is obvious from the metallicity profile (Figs. 6, 7, and 8).

Finally, it should be pointed out that our present discussions in this and previous sections depend heavily on the outcome of retrieval analysis, i.e., the value of X/H. In the literature, such a value is not necessarily consistent among different studies for the same target. For instance, very different values are reported for GJ 436 b (Morley et al. 2017) and HAT-P-11

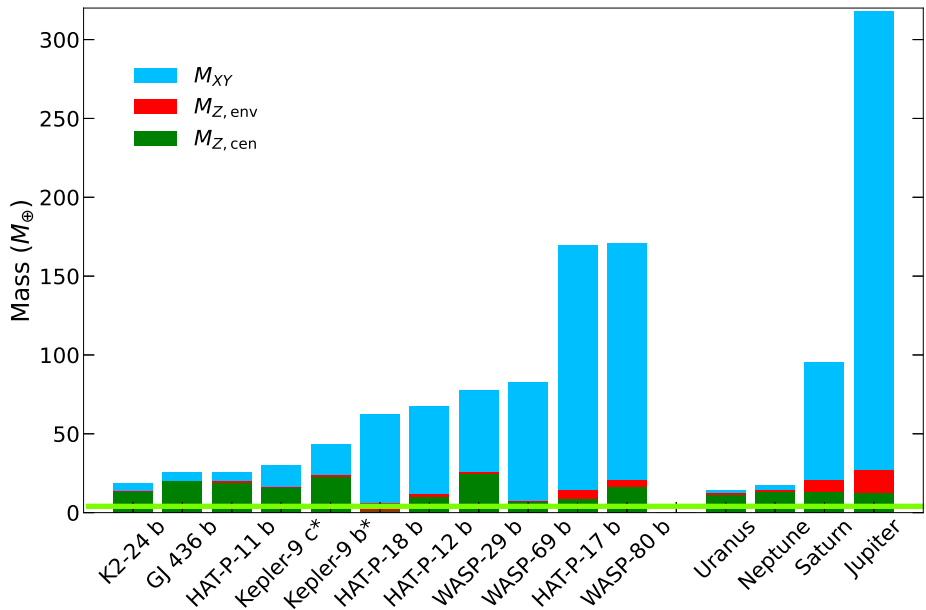

**Fig. 5** The mass budget of solar and extrasolar planets. The gas mass ($M_{XY}$), the solid mass in the envelope ($M_{Z,\text{env}}$), and the solid mass in the planet center ($M_{Z,\text{cen}}$) are denoted by the blue, red, and green bars, respectively. Heavy elements in envelopes tend to increase with increasing planet mass, but some measurements suggest that most of them very likely concentrate in the central region currently. The horizontal light green line represents a predicted core mass of $4M_\oplus$ (Hasegawa et al. 2018)

b (Chachan et al. 2019). This is clearly seen in Fig. 9. The origin of such big differences remains a topic to be studied by the community (e.g., Barstow et al. 2020; Barstow 2020; Welbanks and Madhusudhan 2022; Barstow et al. 2022).

It is clear that if high-metallicity solutions would be reasonable, then the well-mixed assumption becomes valid and the transit-derived metallicity becomes comparable to the density-derived metallicity. On the other hand, if low-metallicity solutions would be reasonable, then the well-mixed assumption would not be justified and heavy elements would concentrate in the central region of planets. The range of different findings for the transit-derived mass metallicity relationship (discussed in Sect. 3.2) prevent a firm, obervationally-motivated conclusion about the presence or absence of composition gradients at the present time. Thus, obtaining consistent results from different studies would be a critical next step for the community to take, in order to derive a meaningful understanding of exoplanet atmospheres and their evolution from transit observations.

### 4.5 Implications for Heat Flow and Magnetic Fields

Composition gradients in the deep envelope are key to understanding heat transport from the interior and the location of fully convective regions; the implications of this are wide-ranging and include topics such as heat flow for core-powered mass loss in sub-Neptunes (Misener and Schlichting 2022), the generation of non-dipole magnetic field components in Uranus and Neptune (Stanley and Bloxham 2004), and the interior structure of Jupiter and Saturn (Mankovich and Fortney 2020). Because the majority of the heavy elements are likely to be in the envelope (Thorngren et al. 2016), it is essential to explore the possibility of strong radial composition gradients. When applied at a population level, it is possible that sufficiently precise measurements of exoplanet metallicity may even provide constraints that are useful for models of double diffusive convection (Leconte and Chabrier 2012) applied in an exoplanet context.

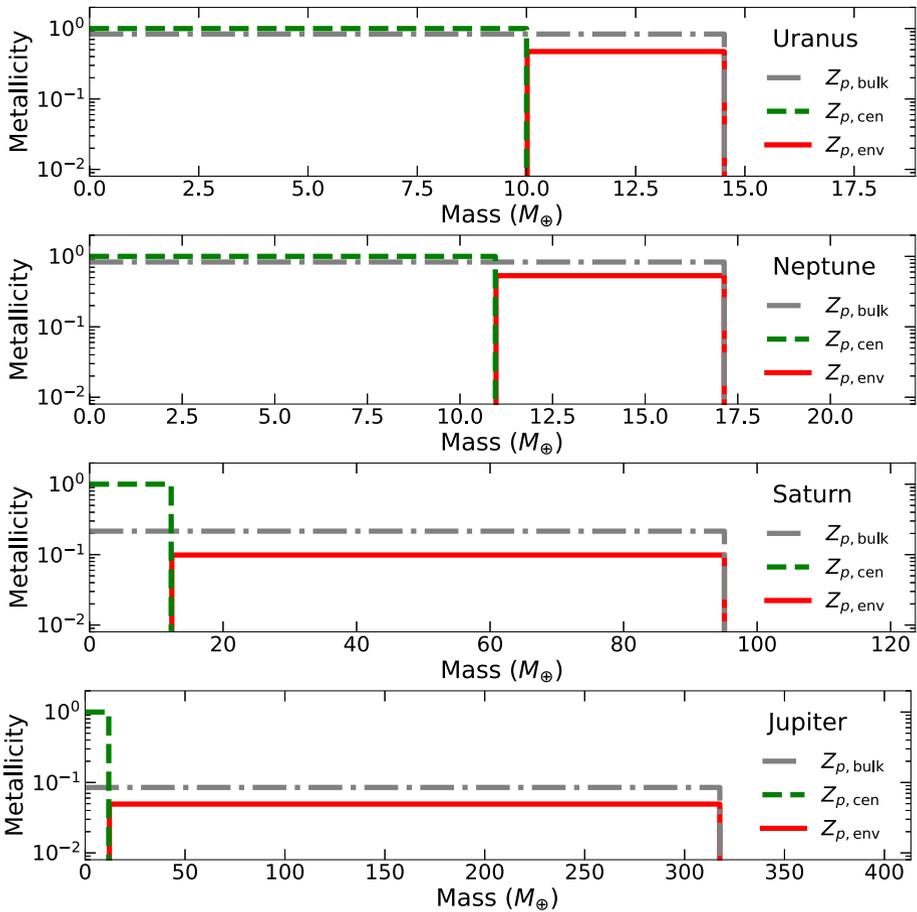

**Fig. 6** The metallicity profile of giant planets in the solar system. The entirely well-mixed case corresponds to the profile of $Z_{p,\text{bulk}}$ (the gray dashed-dotted lines), while the globally well-mixed case is represented by the profiles of $Z_{p,\text{cen}}$ and $Z_{p,\text{env}}$ (the green dashed and the red solid lines, respectively)

### 4.6 Implications for Planet Formation Processes

The possibility of sequestering metals in the core has been considered for exoplanets. However, the question of the initial core mass remains, and the bare cores of exoplanets that are believed to have once been sub-Neptunes may offer some insight. Using a sample of small exoplanets, the core mass of evaporated sub-Neptunes has been estimated as $4.8 \pm 1.8 M_\oplus$ (Swain et al. 2019), consistent with theoretical predictions of $4.3 \pm 1.3 M_\oplus$ (Lee 2019), and the core mass of sub-Neptunes that retain part of the original primordial envelope has been estimated as $6.5 \pm 2.9 M_\oplus$ (Estrela et al. 2020). Thus, under the plausible assumption that the cores of Jupiter-mass planets should be at least as massive than the cores of sub-Neptunes, we would estimate the core mass is at least $\sim 5 M_\oplus$, consistent with planet formation models (e.g., Mordasini et al. 2014; Hasegawa and Pudritz 2014).

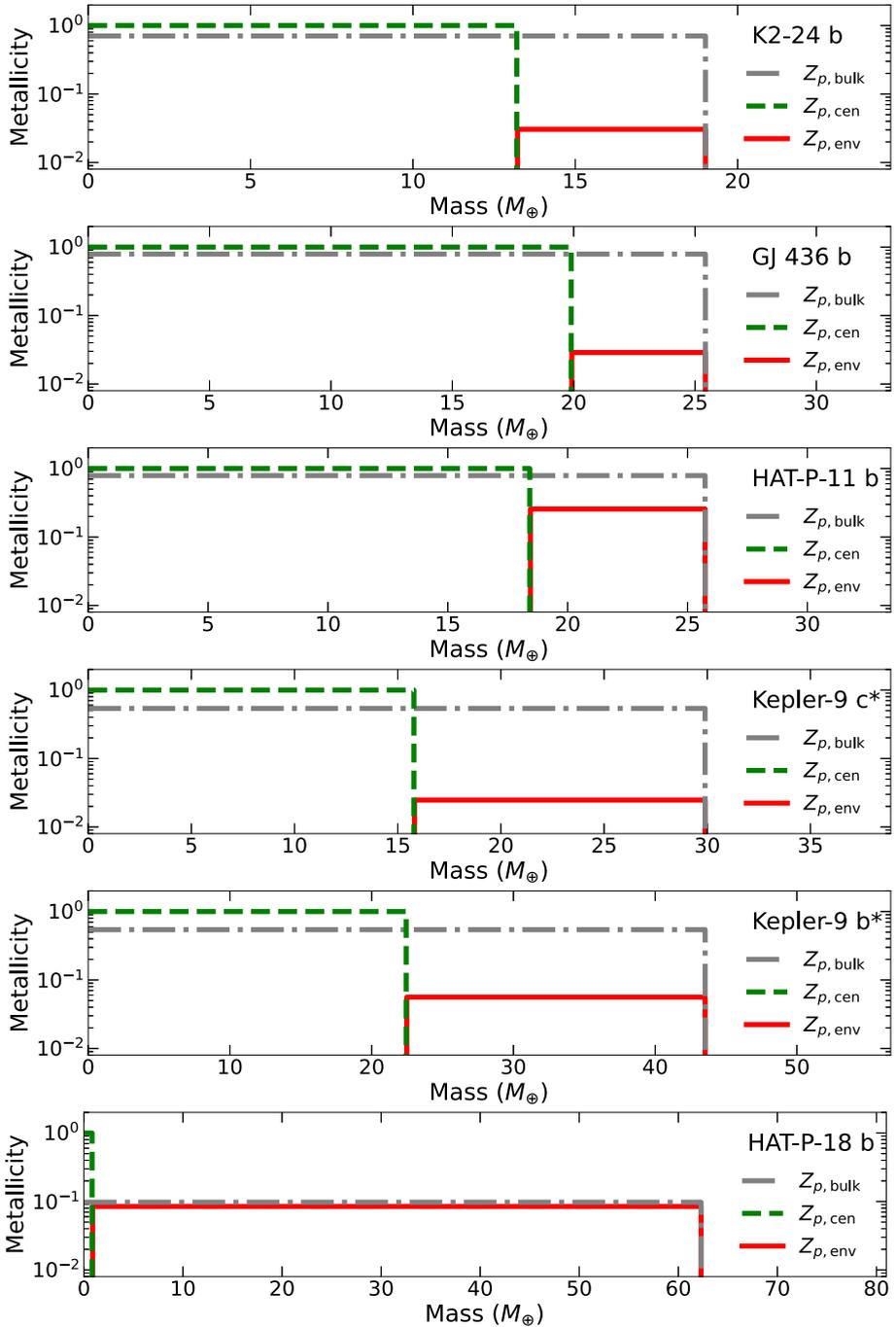

**Fig. 7** The metallicity profile of extrasolar planets as in Fig. 6

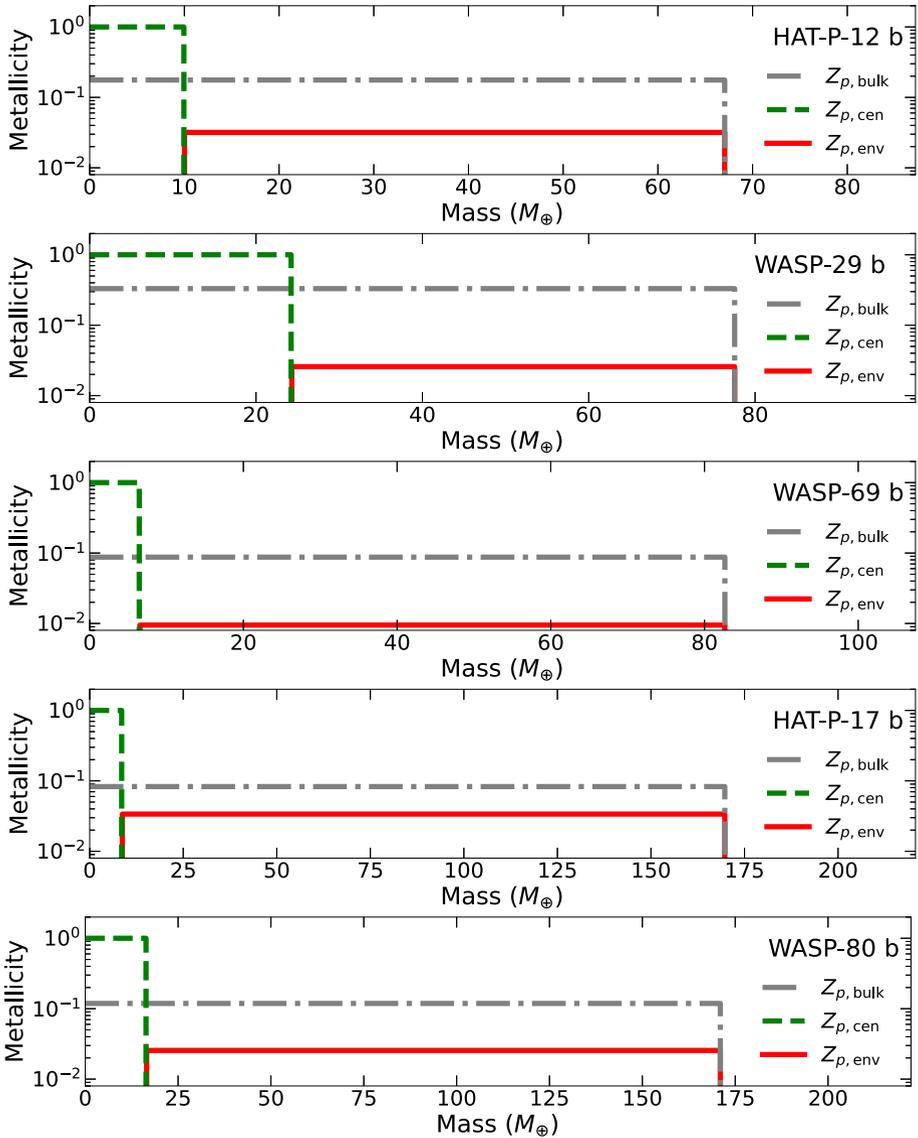

**Fig. 8** The metallicity profile of extrasolar planets as in Fig. 6 (continued)

Stochastic variations in the gas-to-solid ratio during the core-formation and envelope-assembly process may provide "scatter" around the bulk density–derived mass-metallicity trend, as found in simulations (Mordasini et al. 2016); however, more extreme variations in the bulk density–determined planet metallicity could be possible. The high metal content of a subset of hot Jupiters has been noted (Leconte et al. 2009), and it is thought that the metal enrichment accompanying the envelope-assembly phase could be enhanced through migration (e.g., Alibert et al. 2005a). The analysis by Thorngren et al. (2016) of the fractional metal content is conservative in that the total planetary metal content increases even more if all the metals are placed in the core.

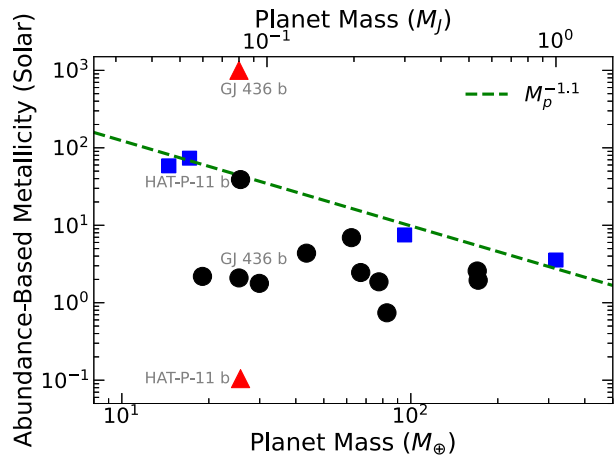

**Fig. 9** The outcome of retrieval analyses available in the literature is not necessarily consistent for the same target. The abundance-based metallicity derived for GJ 436 b and HAT-P-11 b show huge differences for different studies. As examples, exoplanets covered by both Thorngren et al. (2016) and Edwards et al. (2023) are plotted. Also see Fig. 10. The fitting result of Kreidberg et al. (2014b) is shown by the green dashed line

## 5 Summary and Conclusions

A recurring theme in exoplanet studies is the comparison of an exoplanet mass-metallicity relationship with the "Solar System trend"—which is based on the C/H ratio of Solar System planets. There are, however, several difficulties with this approach that deserve further investigation. Firstly, we lack evidence that carbon tracks the bulk metallicity in all solar system planets (indeed, there is evidence to the contrary). Secondly, the metallicity of Uranus and Neptune may be correlated more with formation location than with planet mass. Thirdly, all the gas and ice giants in our solar system have evidence for composition gradients, and in some cases, these are sufficiently strong to suppress convection in some regions. The magnetic field structure and interior heat flux may be directly impacted by suppressing convection.

However, the question of composition gradients remains open for exoplanets. Due to the difference in regions of the envelope that are probed, insight into the likely presence, or absence, of composition gradients could be provided by comparing the bulk density–derived mass-metallicity relationship to the transit-derived mass-metallicity relationship.

The bulk density–derived mass metallicity seen in exoplanets is similar to the mass-to-carbon abundance seen in solar system giant planets. To date, attempts to determine the exoplanet mass-metallicity relationship from transit spectroscopy measurements of exoplanets have relied mostly on Hubble WFC3 measurements, although this is changing with the advent of JWST. When compared with the bulk density–derived mass-metallicity relationship, transit-derived mass-metallicity relationships are varied with examples of studies supporting both a well-mixed interpretation and a composition gradient interpretation. Fortunately, the availability of JWST and, in the future, Ariel (Tinetti et al. 2018) will provide the wavelength coverage needed to provide robust estimates of exoplanet atmospheric metallicity and, when compared with bulk density–derived mass-metallicity estimates, provide insights into the degree to which exoplanets support composition gradients similar to what we find in the solar system.

## Appendix A: Conversion of the Abundance-Based, Transit-Derived Metallicity to Mass-Based Metallicity

Transit observations coupled with retrieval analysis provide the enhancement/reduction factor of elements residing in exoplanet atmospheres relative to host stellar metallicity. Here, we present a brief summary of how the abundance-based, transit-derived metallicity is converted to the mass-based one.

Suppose that transit observations infer that exoplanet atmospheres exhibit the abundance enhancement/reduction (X/H) of an element (X) relative to solar abundance. Then, the mass fraction ($m_X$) of the element that has $A_X$ nucleons is written as

$$m_X = A_X N_X \frac{X}{H}, \tag{A1}$$

where $N_X$ is the number density of the element at solar metallicity and $m_X$ is normalized by the atomic mass unit.

When the C/O ratio is also known, which tends to be common for characterization of exoplanet atmospheres, the total number ($N_{tot}$) of elements in atmospheres is written as

$$N_{tot} = \sum_X f_{C/O} N_X \frac{X}{H}, \tag{A2}$$

where it is assumed that X/H = 1 for hydrogen and helium, and the C/O ratio is taken into account as follows:

$$f_{C/O} = \begin{cases} 1 & (X \neq C) \\ C/O & (X = C). \end{cases}$$

Note that the solar value is adopted for $N_{He}$, and $N_H$ is computed such that $N_{tot}$ becomes unity.

Thus, the mass-based, transit-derived metallicity is computed as

$$Z^{ob}_{p,env} \equiv \sum_{X \neq H, He} f_{C/O} A_X N_X \frac{X}{H} \Big/ \sum_X f_{C/O} A_X N_X \frac{X}{H}. \tag{A3}$$

In general, different elements would have different values of X/H. However, current observations still do not allow one to reliably derive such values for all the elements. We therefore assume that one value of X/H would be applicable to all the elements in our calculations presented in the main text. We also adopt the value of $N_X$ from Asplund et al. (2009).

Figure 10 shows the abundance-based metallicity and the mass-based one on the left and right panels, respectively. As examples, we focus on targets that are both covered by Thorngren et al. (2016) and Edwards et al. (2023). The mass-based metallicity is computed from the abundance-based one, as described above. One may notice that compared with the abundance-based one, the mass-based metallicity tends to exhibit large scatter, leading to a slightly shallow correlation with planet mass. However, the overall trend is comparable between the abundance-based and mass-based metallicities.

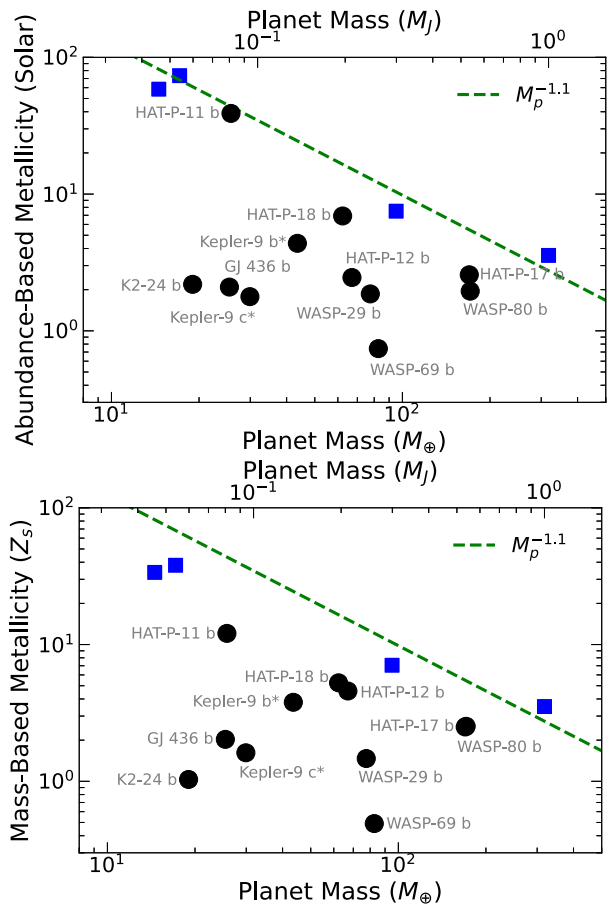

**Fig. 10** The abundance-based and the mass-based metallicities as in Fig. 9. Exoplanets studied in both Thorngren et al. (2016) and Edwards et al. (2023) are chosen in this plot. For comparison purpose, solar system planets are included. The mass-based metallicity may show large scatter and a slightly shallow correlation with planet mass

**Acknowledgements** This research has made use of the NASA Exoplanet Archive, which is operated by the California Institute of Technology, under contract with NASA under the Exoplanet Exploration Program. This research was carried out at the Jet Propulsion Laboratory, California Institute of Technology, under a contract with NASA. YH thanks Artem Aguichine and Lorenzo Mugnai for stimulating discussions and the support from the NASA Exoplanets Research Program through grant 20-XRP20_2-0008. The research was carried out at the Jet Propulsion Laboratory, California Institute of Technology, under a contract with the National Aeronautics and Space Administration (80NM0018D0004). © all rights reserved.

**Author Contribution** MS led the writing, YH developed the formalism laid out in Sect. 4.5 and contributed to the analysis, DT provided extensive input on the state of interior modeling, GR performed simulations used in the preparation of the manuscript.

**Funding** See acknowledgements.

**Availability of data and materials** Not applicable

**Code Availability** Not applicable

## Declarations

**Consent for publication** The authors all consent for this work to be published.

**Competing Interests** The authors declare that they have no conflicts of interest or competing interests in undertaking this work.

## Authors and Affiliations

**Mark R. Swain[1] · Yasuhiro Hasegawa[1] · Daniel P. Thorngren[2] · Gaël M. Roudier[1]**

✉ M.R. Swain
mark.r.swain@jpl.nasa.gov

Y. Hasegawa
yasuhiro.hasegawa@jpl.nasa.gov

D.P. Thorngren
dpthorngren@gmail.com

G.M. Roudier
gael.m.roudier@jpl.nasa.gov

[1] Astrophysics Section, Jet Propulsion Laboratory, California Institute of Technology, 4800 Oak Grove Drive, Pasadena, 91109, CA, USA

[2] Department of Physics & Astronomy, Johns Hopkins University, 3400 N. Charles St., Baltimore, 21210, MD, USA